\documentclass[]{aa}
\usepackage[dvips]{graphicx}
\usepackage{natbib}
\bibpunct{(}{)}{;}{a}{}{,}

\def\kms{km~s$^{-1}$}
\def\Teff{$T_{\rm eff}$}

\def\sch{Schwarzschild}

\def\p1{Paper~I}

\begin{document}

%\thesaurus{06(          % A&A Section 6: Form. struct. and evolut. of stars
%              08.16.4;  % Stars: AGB and post-AGB
%              08.01.3;  % Stars: atmospheremots-cles
%              08.12.1;  % Stars: late-type
%              08.15.1;  % Stars: oscillations
%              08.22.3;  % Stars: variables: general
%              02.19.1)  % Shock waves
%           }

\title{Envelope tomography of long-period variable stars\thanks{Based on 
observations made at Observatoire de Haute Provence, operated by the Centre 
National de la Recherche Scientifique, France}}
\subtitle{II. Method}
 
\author{Rodrigo Alvarez\inst{1}  
   \and Alain Jorissen\inst{1}\fnmsep\thanks{Research Associate, F.N.R.S. (Belgium)}
   \and Bertrand Plez\inst{2}
   \and Denis Gillet\inst{3}
   \and Andr\'e Fokin\inst{4}
   \and Maya Dedecker\inst{1}\fnmsep\thanks{F.R.I.A. Research Assistant (Belgium)}
}
\institute{Institut d'Astronomie et d'Astrophysique, 
           Universit\'e Libre de Bruxelles, 
           C.P.\,226, Boulevard du Triomphe,
           1050 Bruxelles, Belgium
           (ralvarez, ajorisse, dedecker@astro.ulb.ac.be)
\and
           GRAAL,
           Universit\'e Montpellier II,
           cc072,
           34095 Montpellier cedex 05, France
           (plez@graal.univ-montp2.fr)
\and
           Observatoire de Haute-Provence,
           04870 Saint-Michel l'Observatoire, France
           (gillet@obs-hp.fr)
\and
           Institute for Astronomy of the Russia Academy of Sciences,
           48 Pjatnitskaja,
           109017 Moscow, Russia
           (fokin@inasan.rssi.ru)
}

\offprints{A.\ Jorissen \email{ajorisse@astro.ulb.ac.be}}

\date{Received date / Accepted date}

\titlerunning{Envelope tomography of LPV stars. II}
\authorrunning{R.\ Alvarez et al.}

\abstract{
A tomographic method is described that makes it possible to follow the
propagation of shock waves across the photosphere of long-period
variable stars. The method relies on the
correlation of the observed spectrum with 
numerical masks probing layers of different atmospheric  depths. 
The formation depth of spectral lines is derived 
from synthetic spectra of non-variable 
red giant stars. When applied to
Mira stars around maximum light, the tomographic masks
reveal that the deepest photospheric layers are
generally characterized by blueshifted absorption lines (attesting their
upward motion), whereas the uppermost layers generally exhibit redshifted
absorption lines (indicating their infalling motion). 
Double absorption lines 
are found in intermediate layers, where the shock front is located. At later
phases, the shock front is  seen moving towards upper layers, until it leaves
the photosphere.
}

\maketitle

\section{Introduction}

Besides the brightness fluctuations, long-period variable stars
(LPVs) also exhibit striking 
spectral changes, like 
the doubling of several absorption lines around maximum light, first
reported by Adams (1941).
The line-doubling phenomenon has been source of conflicting theories. 
\sch\ (1952) has suggested (originally in relation with W Vir Cepheids) that 
the doubling of the absorption lines around maximum light is related
to the passage of a shock wave through the photosphere.  
Alternative models accounting for line doubling without 
resorting to differential atmospheric motions have been proposed by
Karp (1975) and  Gillet et al. (1985).

In \p1\ of this series (Alvarez et al. 2000), 
it was shown that the temporal evolution of the red and blue 
peaks of the double absorption lines of the Mira variable RT\,Cyg 
follows the ``\sch\ scenario''. This result thus indicates that the 
line-doubling phenomenon is caused by a shock wave propagating in the 
photosphere. In \p1, it was shown that this propagation may even be
followed by resorting to 
a tomographic technique. The
method relies 
on the correlation of the observed spectrum with 
numerical masks probing layers of different atmospheric  depths. This 
technique applied to RT~Cyg revealed that the evolution of the line doubling
{\it with atmospheric depth} is also consistent with the 
\sch\ scenario.

This paper presents the implementation of this tomographic technique. 
Its application to 
a large sample of LPV stars is presented in Paper~III.  
The possible relationship between absorption 
line-doubling and emission lines will be investigated in a forthcoming paper 
in this series.

\section{Scanning different parts of the atmosphere with the default ELODIE 
correlation masks}
\label{Sect:templates}

A  monitoring of a large sample of LPV 
stars was performed with the fibre-fed echelle spectrograph ELODIE
(Baranne et al.\ 1996) at the Observatoire de Haute-Provence
(France). 
The observations and the star sample are 
described in detail in Paper~III (Tables~1 and 2). 
The
ELODIE spectrograph 
is designed to perform very accurate radial-velocity measurements 
by cross-correlating the stellar spectrum with numerical masks.
In a first step, the cross-correlation functions (CCF) 
of the program stars were computed using two different  
numerical masks included in the ELODIE reduction
software:
(i) a K0\,III mask constructed by Baranne et al.\ (1996);
(ii) a M4\,V mask constructed by Delfosse et al.\ (1999) from an ELODIE 
spectrum of Barnard's star (Gl~699) applying the method of Baranne et 
al.\ (1979).

\subsection{Comparison of the CCFs obtained with the K0- and M4-templates}
\label{Sect:compar_ccf}

\begin{figure}
%  \resizebox{\hsize}{!}{\includegraphics{vumi.N7.cor.ps}}
  \resizebox{\hsize}{!}{\includegraphics{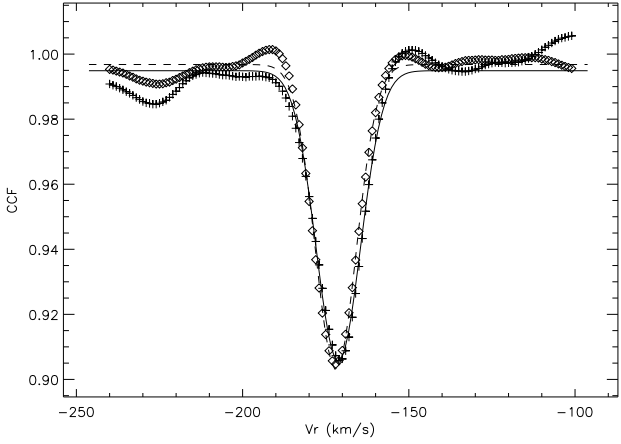}}
  \caption[]{Cross-correlation profile of V\,UMi obtained with the default 
  K0\,III (crosses) and M4\,V (diamonds) masks, during night N7 
  at phase $\sim$0.99 (the night number refers to Table~2 of Paper~III); 
the radial velocities as given by gaussian fits
  are: $-$171.2~\kms\ (K0, solid line) and $-$171.9~\kms\ (M4, dashed line)}
  \label{Fig:ccf_vumi_N7}
\end{figure}

\begin{figure}
%  \resizebox{\hsize}{!}{\includegraphics{ruher.N9.cor.ps}}
  \resizebox{\hsize}{!}{\includegraphics{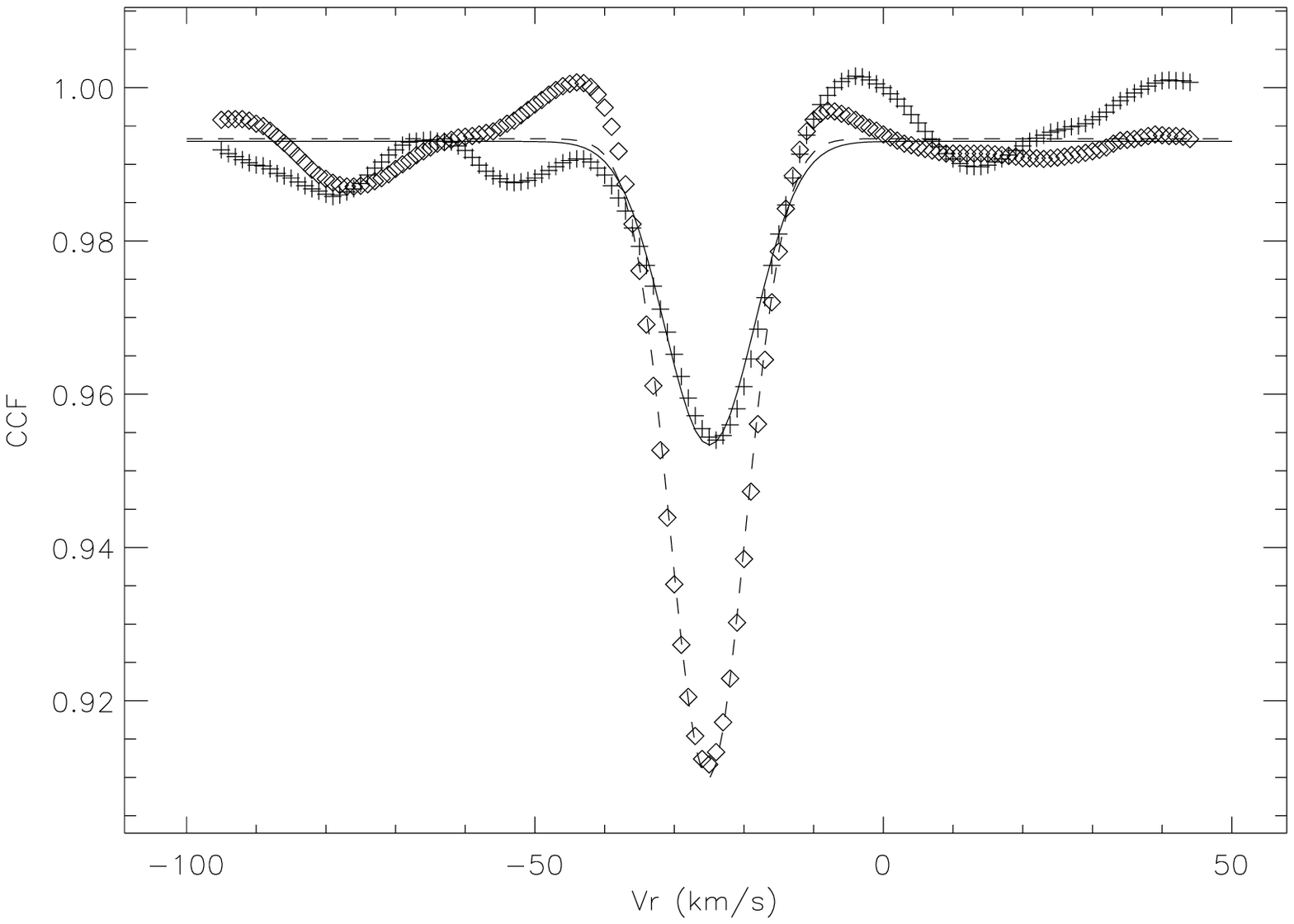}}
  \caption[]{Same as Fig.~\ref{Fig:ccf_vumi_N7} for RU\,Her (night N9, 
  phase 0.02). Radial velocities are: $-$25.0~\kms\ (K0, solid line) and 
  $-$25.2~\kms\ 
  (M4, dashed line)}
  \label{Fig:ccf_ruher_N9}
\end{figure}

\begin{figure}
%  \resizebox{\hsize}{!}{\includegraphics{rtcyg.N2.cor.ps}}
  \resizebox{\hsize}{!}{\includegraphics{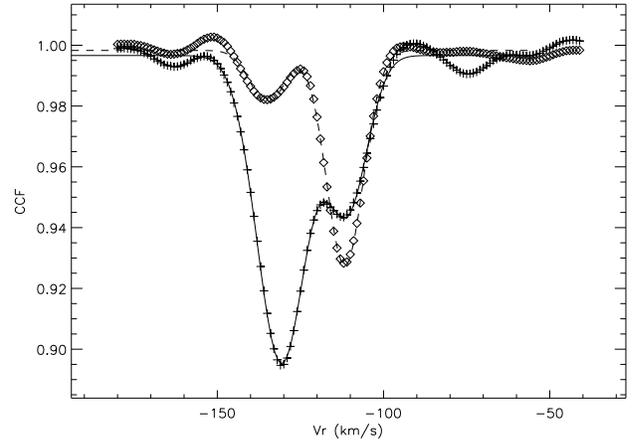}}
  \caption[]{Same as Fig.~\ref{Fig:ccf_vumi_N7} for RT\,Cyg (night N2, 
  phase 1.10). Radial velocities are: $-$130.7/$-$111.0~\kms\ (K0, solid 
  line) and $-$134.9/$-$111.6~\kms\ (M4, dashed line)}
  \label{Fig:ccf_rtcyg_N2}
\end{figure}

\begin{figure}
%  \resizebox{\hsize}{!}{\includegraphics{uzhya.N7.cor.ps}}
  \resizebox{\hsize}{!}{\includegraphics{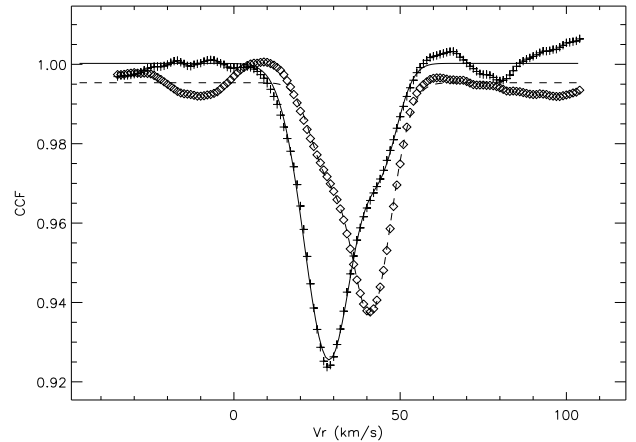}}
  \caption[]{Same as Fig.~\ref{Fig:ccf_vumi_N7} for UZ\,Hya (night N7, 
  phase 1.09). Radial velocities are: +28.6/+44.8~\kms\ (K0, solid line) 
  and +27.4/+41.0~\kms\ 
  (M4, dashed line)}
  \label{Fig:ccf_uzhya_N7}
\end{figure}

\begin{figure}
%  \resizebox{\hsize}{!}{\includegraphics{vtau.N15.cor.ps}}
  \resizebox{\hsize}{!}{\includegraphics{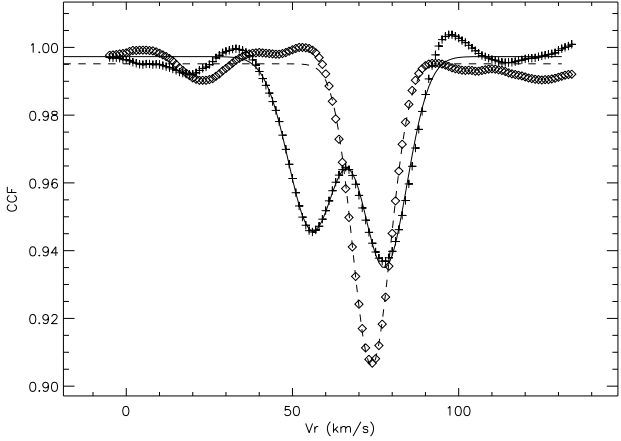}}
  \caption[]{Same as Fig.~\ref{Fig:ccf_vumi_N7} for V\,Tau (night N15, 
  phase 2.01). Radial velocities are: +56.0/+77.8~\kms\ (K0, solid line) 
  and +73.8~\kms\ (M4, dashed line)}
  \label{Fig:ccf_vtau_N15}
\end{figure}

\begin{figure}
%  \resizebox{\hsize}{!}{\includegraphics{rycep.N4.cor.ps}}
  \resizebox{\hsize}{!}{\includegraphics{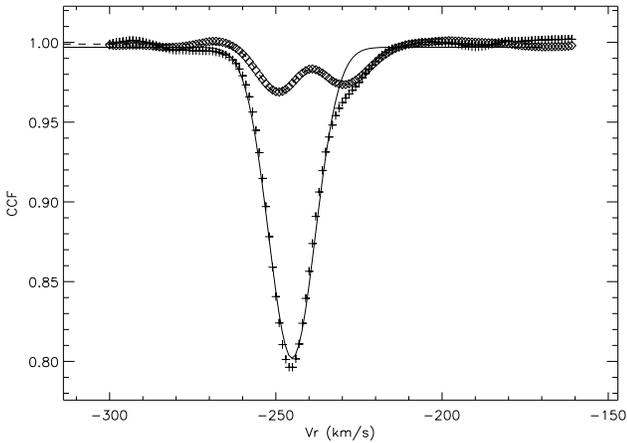}}
  \caption[]{Same as Fig.~\ref{Fig:ccf_vumi_N7} for RY\,Cep (night N4, 
  phase 1.03:). Radial velocities are: $-$245.1~\kms\ (K0, solid line) 
  and $-$249.5/$-$229.3~\kms\ (M4, dashed line)}
  \label{Fig:ccf_rycep_N4}
\end{figure}

\begin{figure}
%  \resizebox{\hsize}{!}{\includegraphics{rpeg.N15.cor.ps}}
  \resizebox{\hsize}{!}{\includegraphics{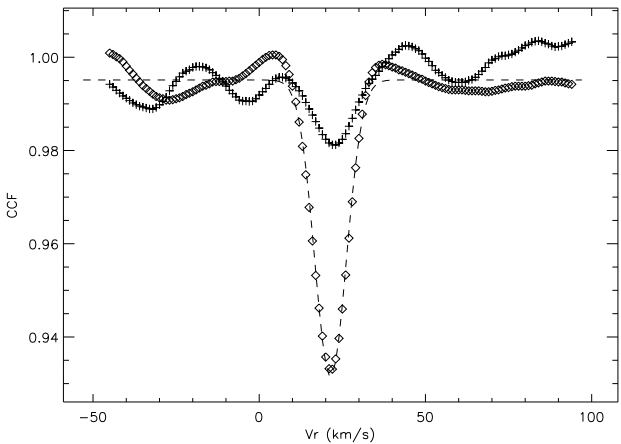}}
  \caption[]{Same as Fig.~\ref{Fig:ccf_vumi_N7} for R\,Peg (night N15, 
  phase 2.00). Radial velocity is: +21.5~\kms\ (M4, dashed line)}
  \label{Fig:ccf_rpeg_N15}
\end{figure}

\begin{figure}
%  \resizebox{\hsize}{!}{\includegraphics{tcep.N1.cor.ps}}
  \resizebox{\hsize}{!}{\includegraphics{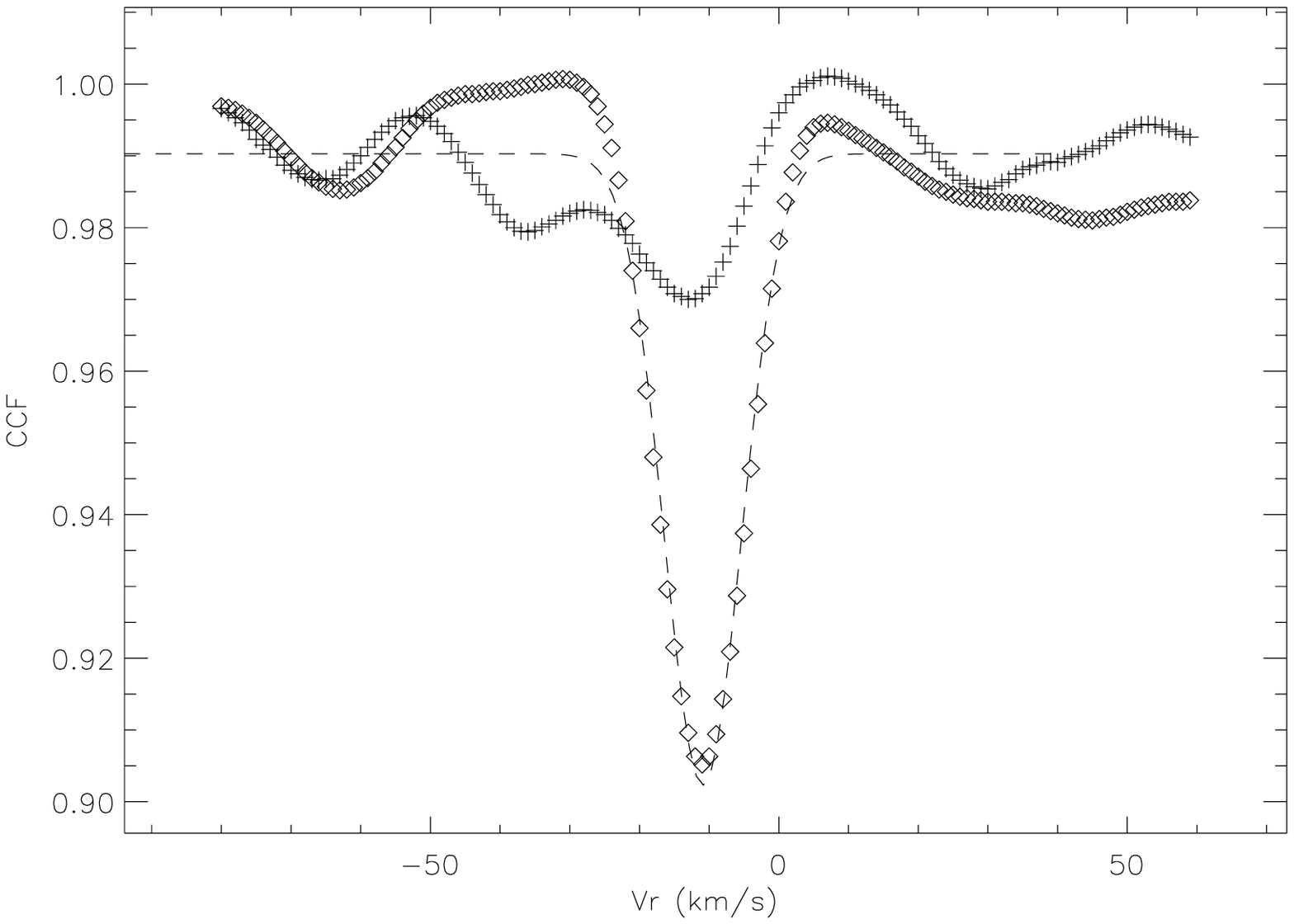}}
  \caption[]{Same as Fig.~\ref{Fig:ccf_vumi_N7} for T\,Cep (night N1, 
  phase 0.23). Radial velocity is: $-$10.9~\kms\ (M4, dashed line)}
  \label{Fig:ccf_tcep_N1}
\end{figure}

\begin{figure}
%  \resizebox{\hsize}{!}{\includegraphics{rtdra.N12.cor.ps}}
  \resizebox{\hsize}{!}{\includegraphics{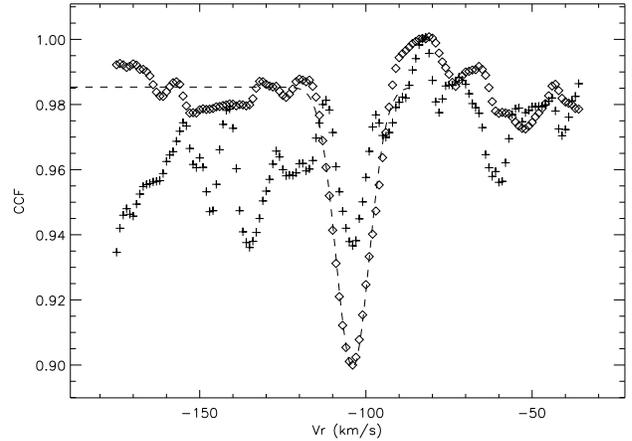}}
  \caption[]{Same as Fig.~\ref{Fig:ccf_vumi_N7} for RT\,Dra (night N12, 
  phase 1.13). Radial velocity is: $-$104.0~\kms\ (M4, dashed line)}
  \label{Fig:ccf_rtdra_N12}
\end{figure}

The CCFs obtained for a given LPV star with the K0- and 
M4-templates are often quite different. Almost all situations may be 
encountered. The most common are:
(i) single peak with similar or different contrasts (Figs.~\ref{Fig:ccf_vumi_N7}
and \ref{Fig:ccf_ruher_N9},
respectively);
(ii) double or asymmetrical peak with both templates
(Figs.~\ref{Fig:ccf_rtcyg_N2} and \ref{Fig:ccf_uzhya_N7});
(iii) double peak with the K0-template, red peak with the M4-template
(Fig.~\ref{Fig:ccf_vtau_N15});
(iv) blue peak with the K0-template, double peak with the M4-template
(Fig.~\ref{Fig:ccf_rycep_N4});
(v) doubtful single peak (Fig.~\ref{Fig:ccf_rpeg_N15}), doubtful double peak 
(Fig.~\ref{Fig:ccf_tcep_N1}) or noisy profile (Fig.~\ref{Fig:ccf_rtdra_N12}) 
with the K0-template, single peak with the M4-template.

% Figure CCF V UMi, N7, K0 & M4 -> Fig:ccf_vumi_N7

% Figure CCF RU Her, N9, K0 & M4-> Fig:ccf_ruher_N9

% Figure CCF RT Cyg, N2, K0 & M4-> Fig:ccf_rtcyg_N2

% Figure CCF UZ Hya, N2, K0 & M4-> Fig:ccf_uzhya_N7

% Figure CCF V Tau, N15, K0 & M4-> Fig:ccf_vtau_N15

% Figure CCF RY Cep, N4, K0 & M4-> Fig:ccf_rycep_N4

% Figure CCF R Peg, N15, K0 & M4-> Fig:ccf_rpeg_N15

% Figure CCF T Cep, N1, K0 & M4-> Fig:ccf_tcep_N1

% Figure CCF RT Dra, N12, K0 & M4-> Fig:ccf_rtdra_N12

\subsection{Interpretation: a first step towards tomography}
\label{Sect:1_step}

\begin{figure}
% \resizebox{\hsize}{!}{\includegraphics{mgem_templates.ps}}
 \resizebox{\hsize}{!}{\includegraphics{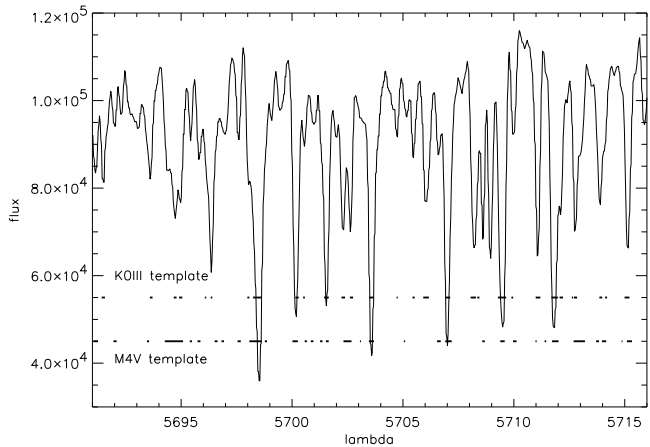}}
 \caption[]{Part of the spectrum of $\mu$\,Gem (M3), observed during night 
 N6. The thick segments represent the ``holes'' of the K0- and M4-templates}
 \label{Fig:mugem_templates}
\end{figure}

The variety of CCF pairs obtained for a given star with the K0- and
M4-templates (Sect.~\ref{Sect:compar_ccf}) may be easily explained if one
admits that the K0-template probes deeper layers than the M4-template (the
validity of this assumption will in fact be demonstrated in
Sect.~\ref{Sect:var}). As
may be seen on Fig.~\ref{Fig:mugem_templates}, the two templates catch different
groups of spectral lines. Depending on the
phase and on the spectral and chemical types, both, none or only one of
the default templates will be able to detect the velocity discontinuity
between the deeper, ascending layers and the upper, infalling layers.

% Figure mu Gem and the default templates -> Fig:mugem_templates

It will be shown in Paper~III (Table~3 and Sect.~4.3.1) that double peaks 
are observed more often with the K0-template, whereas the M4-template
often reveals only a single red peak (as in Fig.~\ref{Fig:ccf_vtau_N15}).
This conclusion is a natural consequence of the fact that
(i) the M4-template is more sensitive to the cooler, outermost layers
which generally exhibit an infall motion around maximum light  
(hence the corresponding CCF is a red peak),
(ii) the K0-template is more sensitive to hotter, i.e.\ inner, layers and,
hence, is more likely to catch the discontinuity between the ascending
and infalling layers (corresponding to a double-peak CCF). A
  similar conclusion was reached by Hinkle et al. (1984), since the
   CO $\Delta v = 3$ lines 1.6 $\mu$m, forming deeper in the photospere
  than the blue-violet optical lines, exhibit line doubling
  while the latter do not.  

Sometimes however, the M4-template is also able to detect a double 
peak. In these cases, the K0-template may reveal either a double peak as well, 
the blue peak being then more pronounced than with the M4-template
(Figs.~\ref{Fig:ccf_rtcyg_N2} and \ref{Fig:ccf_uzhya_N7}), or solely the
blue peak (Fig.~\ref{Fig:ccf_rycep_N4}).

When the star is at minimum light, the discontinuity front has left the
photosphere so that the line doubling is no more visible. The late spectral
types exhibited by most LPVs at minimum light prevent us from obtaining
a well-defined CCF with the K0-template, which becomes useless. On the 
contrary, a single-peak CCF can still be obtained with the M4-template
(Figs.~\ref{Fig:ccf_tcep_N1} and 
\ref{Fig:ccf_rtdra_N12}) which thus delivers a well-defined CCF 
over the complete light cycle of (oxygen-rich) LPVs. 

\begin{figure}
% \resizebox{\hsize}{!}{\rotatebox{90}{\includegraphics{rtcyg_suivi.ps}}}
 \resizebox{\hsize}{!}{\rotatebox{90}{\includegraphics{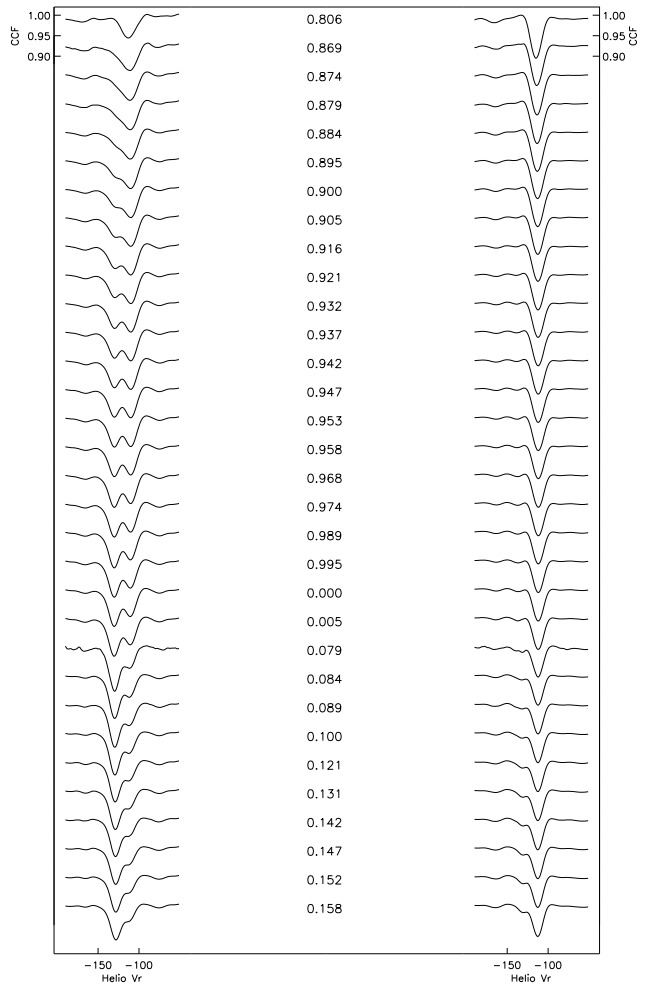}}}
 \caption[]{Sequence of CCFs for RT\,Cyg obtained with the K0\,III mask 
 (left side) and the M4\,V mask (right side) during a two-month-long 
 monitoring in August/September 1999. The labels denote the visual phases}
 \label{Fig:rtcyg_survey}
\end{figure}

\begin{figure}
%  \resizebox{15cm}{20cm}{\rotatebox{90}{\includegraphics{rycep_suivi.ps}}}
  \resizebox{8cm}{!}{\rotatebox{90}{\includegraphics{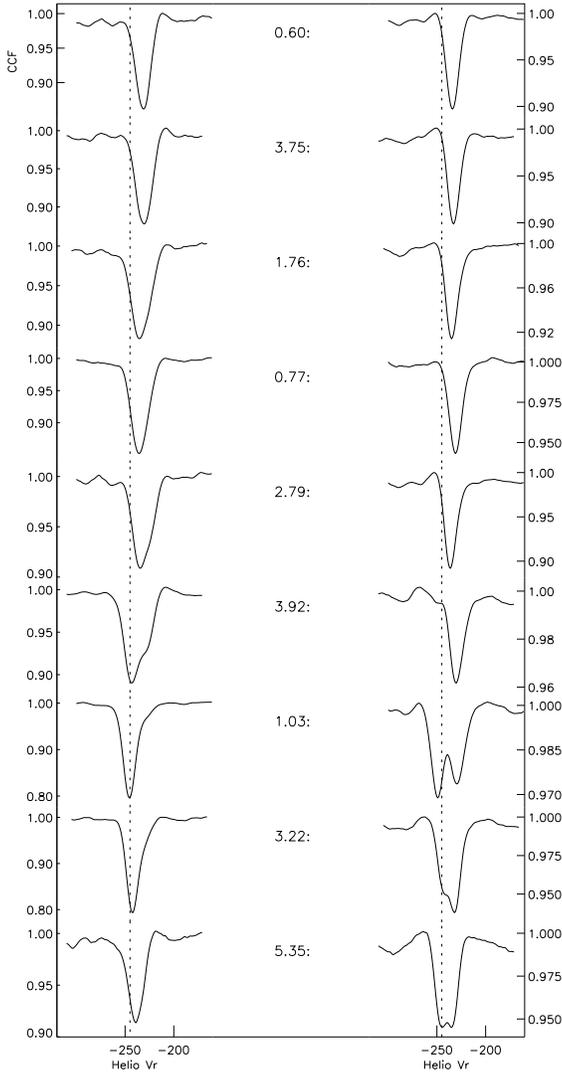}}}
  \caption[]{
  \label{Fig:rycep_survey}
Sequence of CCFs for RY\,Cep with the default K0\,III mask
  (left side) and the M4\,V mask (right side). The labels denote the visual 
  phases. They are ordered according to increasing fractional phase within each
cycle, and are centered around maximum light. The vertical lines are a guide to
the eye}

\end{figure}

Another interesting conclusion may be drawn from the comparison of the
CCFs obtained with the K0- and M4-templates for stars which exhibit double 
peaks: the double peak appears at later phases with the M4-template than 
with the K0-template. As Fig.~\ref{Fig:rtcyg_survey} indicates, a weak secondary
peak appears in RT\,Cyg  at phase 0.10 when observed with the 
M4-template as compared to phase 0.88 with the K0-template. The same
conclusion is reached for RY\,Cep, a Ke-M0e Mira 
(Fig.~\ref{Fig:rycep_survey}).

This phase lag can be interpreted once again in terms of the \sch\ scenario:
since the shock wave propagates outwards and the M4-template scans higher
layers than the K0-template, such a phase shift is not surprising. It 
explains why the doubling is often observed {\it after} maximum light 
with the M4-template.

For the same reason, the red peaks observed with the M4-template are almost
systematically blue-shifted by 2--4 \kms\ as compared to the 
ones observed with the K0-template (see Figs.~\ref{Fig:ccf_uzhya_N7} and
\ref{Fig:ccf_vtau_N15}, and Table~3 of Paper~III). 
This difference may be interpreted in 
terms of a velocity gradient: since the infalling layers scanned by the 
M4-template are located above the ones scanned by the K0-template, it seems 
plausible that these distinct layers fall with different velocities, 
reflecting the acceleration due to the gravity.

The results discussed in this section thus show that  
the simultaneous use of the K0- and the
M4-templates provides insights into the velocity field of LPV
atmospheres. Even more interestingly, the capability of the different
templates to probe different layers in the atmosphere, as determined by their
different spectral (i.e.\ temperature) adequacy, provides a hint on how
to develop a method to investigate the propagation of the shock wave in 
LPV atmospheres. This method, which consists in designing specific numerical
templates to perform the ``tomography'' of the atmosphere, is described in
detail in Sect.~\ref{Sect:tomo}.

% Figure survey of RT Cyg with K0 and M4 -> Fig:rtcyg_survey

% Figure survey of RY Cep with K0 and M4 -> Fig:rycep_survey

\section{The tomographic technique}
\label{Sect:tomo}
The basic idea is to design a set of numerical templates that probe 
different atmospheric layers, in order to follow the propagation of the 
shock wave through the atmosphere (a similar technique was applied to BA-type
supergiants by Kaufer et al. 1997). This technique takes advantage of the 
ability of the cross-correlation method to reveal complex profiles despite 
the severe crowding of the optical spectra. The selective sensitivity of the 
default K0- and M4-templates demonstrates the potentiality of the method
(Sect.~\ref{Sect:templates}). 
The crucial point of the tomographic technique is the selection 
of adequate sets of lines forming at a given depth in the atmosphere. 

\subsection{Synthetic spectra}
\label{Sect:synthetic}

The tomographic method rests on our ability to construct reliable synthetic 
spectra of late-type giant stars and to select the right set of spectral 
lines forming at a given depth. We compute synthetic spectra from
static models 
of red giant stars in spherical symmetry 
(Plez et al.\ 1992; Plez 1992) in the spectral range 
3850--6900~\AA. The synthetic spectra are computed at a high resolution 
($\Delta \lambda =$ 0.03~\AA, which corresponds to the binning step of the 
ELODIE spectra) using a set of routines especially designed for cool 
star atmospheric conditions ({\it Turbospectrum}; see Plez et al. 1993). 
The spectral synthesis includes the updated line data for TiO 
(Plez 1998), H$_2$O and VO as detailed in Bessell et al.\ (1998) and 
Alvarez \& Plez (1998), plus CH, C$_2$, CN, NH, OH, MgH, SiH, CaH, ZrO, and
atomic lines 
(Kurucz \& Bell 1995). It results in more than 2.\,10$^7$ lines. The 
spectra are computed with CNO abundances and $^{12}$C/$^{13}$C, 
$^{14}$N/$^{15}$N and $^{16}$O/$^{18}$O isotopic ratios 
(respectively 13, 1250 and 500) typical of red 
giants  (Smith \& Lambert 1990).

\begin{figure}
% \resizebox{\hsize}{!}{\includegraphics{compar_spec.ps}}
 \resizebox{8cm}{!}{\includegraphics{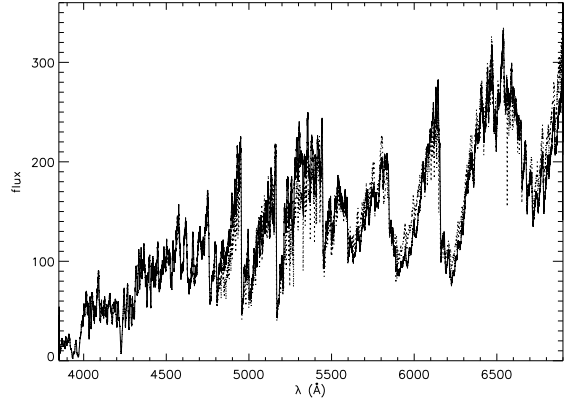}}
 \caption[]{Synthetic spectrum with \Teff =3500~K, log $g$=0.9, 
 $M$=1.5~M$_{\sun}$. The dotted line is part of an optical spectrum of 
 HD\,123657, an M5-giant, from Serote-Roos et al.\ (1996)}
 \label{Fig:compar_spec}
\end{figure}

Figure~\ref{Fig:compar_spec} shows an example of a synthetic spectrum 
(\Teff\ =3500~K). The dotted line is part of an optical spectrum of 
HD\,123657, an M5-giant, obtained by Serote-Roos et al.\ (1996) at a 
resolution of 1.25~\AA. The spectral region covered starts at 4800~\AA. 

% Figure comparison of spectra -> Fig:compar_spec

\subsection{Average depth of formation of spectral lines }
\label{Sect:formation}

The static models described in Sect.~\ref{Sect:synthetic} 
provide us with the monochromatic optical 
depths 
$\tau_{\lambda}$ as a function of geometrical depth, the latter being actually 
represented in terms of  
$x$=$\log \tau_0$, where $\tau_0$ 
is an optical depth at some reference wavelength (here $\lambda_0 =
1.2\;\mu$m). In static models,
the monochromatic optical depth at any reference wavelength is indeed
a monotonic function of the geometrical depth in the atmosphere. 

In the framework of the Eddington-Barbier approximation, the emergent
flux at wavelength $\lambda$ 
is supposed to come from the layer with $\tau_\lambda = 2/3$. 
In the {\it first implementation} of the tomographic method presented here, 
we will assume that the line depression forms as well at that same depth.
This assumption is valid for sufficiently strong lines, 
as shown by Magain (1986).
Using the expression of the contribution
function (CF)\footnote{The contribution function gives the relative 
contribution of the
different atmospheric layers to an observed quantity which can be the
emergent intensity or flux or the line depression in the surface intensity 
or flux} 
to the line depression in intensity, $C_R$,
Magain (1986; see also Gurtovenko et al.\ 1974) has however stressed that 
the region where the line depression is formed may be different 
from the region of origin of the emergent intensity, 
as described by the CF to the line intensity $C_I$
(see Albrow \& Cottrell 1996 for the generalization to the line depression
in flux, in a non-static atmosphere).
Magain demonstrates that $C_R$ is markedly different from  
$C_I$ in the case
of faint lines formed in classical stellar atmospheres (his Fig. 1). 
In the case of a stronger line $C_R$ and $C_I$ are more similar, so
that the line flux and depression originate in the same layers (his Fig. 2).
$C_R$ and $C_I$ are always close to each other in layers above 
$\tau_c = 10^{-2}$ (his Fig. 3), {\it i.e.} for sufficiently strong  lines.
Thus masks probing the deepest
layers are the more likely to yield poor results, as indicated in
Sect.~\ref{Sect:var} (see also Fig.~\ref{Fig:latetypeLPV}).

\subsection{Designing synthetic numerical templates}
\label{Sect:synth_masks}

The synthetic templates are built in two steps. A complete spectrum, with 
all atomic and molecular lines included, is computed. The 
monochromatic optical depth $\tau_{\lambda}$ is calculated at each depth 
of the atmospheric model, for each sampled wavelength. 
The different masks $M_i$ 
are then constructed from the 
collection of $N$   wavelengths $\lambda_{i,j}$ ($1 \le j \le N$) such that 
$x_i \le x(\tau_{\lambda_{i,j}} = 2/3) < x_{i+1} = x_i + \Delta x$, where 
$x$=$\log \tau_0$ is some reference optical depth used to parametrize
the geometrical depth. Each mask $M_i$ thus 
probes lines forming at depths in the range $x_i$, $x_i + \Delta x$.

\begin{figure}
 \resizebox{8cm}{!}{\includegraphics{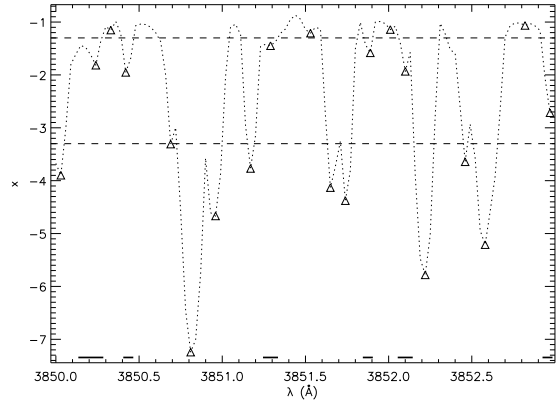}}
 \caption[]{
 \label{Fig:synthmask1.ps}
Illustration of the design of a synthetic template.
 The present template (thick dashes at the bottom of the plot) scans the
 lines forming at depths $x$ such that $-1.3 \le x < -3.3$ (see text)}
\end{figure}

Figure~\ref{Fig:synthmask1.ps} illustrates how the synthetic templates 
are built: for the sake of clarity, only a very small portion (3~\AA) of 
the depth function $x = x(\tau_{\lambda} = 2/3)$ is represented and
the constant 
$\Delta x$ has been assigned a very large value ($\Delta x$=2). The depth
function 
provides the reference optical depth $\tau_0 = 10^x$ -- or, 
equivalently, the geometrical depth --
at which the monochromatic optical depth reaches 2/3.
The depth function $x = x(\tau_{\lambda} = 2/3)$
therefore
identifies the layer from which the monochromatic flux at wavelength
$\lambda_{i,j}$ emerges.  The triangles in Fig.~\ref{Fig:synthmask1.ps} denote
the local minima of the depth function 
(they correspond to the spectral lines): when these minima 
fall in the range scanned 
by a given template (i.e., between the two horizontal dashed lines), a
corresponding ``hole'' is created in this template 
(thick dashes at the bottom of the plot).

The constant $\Delta x$ was chosen after evaluating the contribution function
to the emergent flux for some specific 
atomic lines. 
It appeared that the typical width $W_{\rm CF}$ of the CFs expressed in the 
$x$-scale is about $W_{\rm CF}$=1.5. The Nyquist-Shannon theorem of 
elementary signal theory indicates that the optimal sampling of a given 
function is obtained for $\Delta y$/2 where $\Delta y$ is the ``resolution'' 
of the signal. A value of $\Delta x$=$W_{\rm CF}$/2=0.75 was thus 
adopted to optimize the thickness of the atmospheric region scanned by each 
template $M_i$.

% Figure synthetic mask example 1  -> Fig:synthmask1.ps

\begin{figure}
%  \resizebox{\hsize}{!}{\includegraphics{histo_depthf.ps}}
  \resizebox{8cm}{!}{\includegraphics{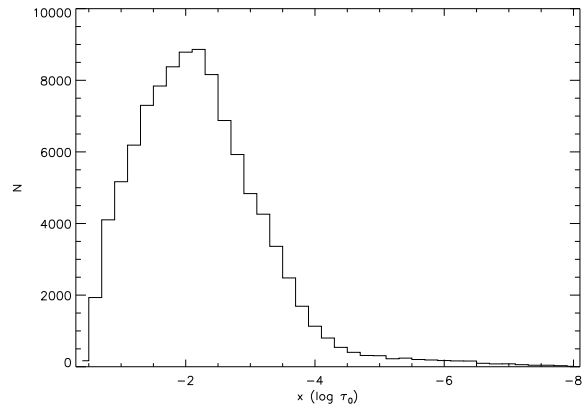}}
  \caption[]{Number of lines as a function of their formation depth 
  (expressed in the $\log \tau_0$ scale; $\tau_0 = 1.2 \mu$m) for a model 
  at \Teff=3500~K}
  % realized with tau_max = -0.3 / tau_min = -8.1
  \label{Fig:h_depthf}
\end{figure}

Three sets of numerical templates were built from models with \Teff=2800,
3500 and 4250~K. Figure~\ref{Fig:h_depthf} shows the 
distribution of the depths of line formation for a model with  \Teff=3500~K:
as expected, the number of lines increases with the atmospheric depth (as
long as $\log \tau_0 \leq -2$). The lines forming in the outer
atmospheric layers are by
far less numerous, but they are also more intense. This justifies to keep 
the same $\Delta x$ for all masks $M_i$: the number of lines and their 
intensities in any given mask vary in opposite directions so as to provide
similar signal-to-noise ratios for the CCFs obtained with the different
masks (see Figs.~\ref{Fig:ccf_mgem_N6_med}, \ref{Fig:ccf_zoph_N1_med} and 
\ref{Fig:ccf_vtau_N4_med}).

% Figure histogram depth function -> Fig:h_depthf

When these masks constructed with the complete set of lines were 
applied to non-variable stars, spurious secondary peaks appeared in the 
CCFs obtained with some of the masks. 
These spurious peaks originate in inaccuracies in the wavelengths of some
molecular transitions, that are difficult to correct.
Therefore, the decision was taken to rely solely on the supposedly more
accurate atomic lines for building the templates. A new set of templates
$M'_i$ were designed following the method described above, except
that they were based on synthetic spectra including only atomic lines.
Since important molecular opacity sources are lacking in these synthetic
spectra, the associated depth functions may no longer be reliable, 
casting doubts on the line assignments leading to the new set of masks $M'_i$.

To circumvent that difficulty, the final masks $C_i$ are constructed by
keeping all the holes from the $M'_i$ templates that appeared as well in the
$M_i$ templates. This way of doing guarantees that only the holes associated
with atomic lines are kept (as they are present in the $M'_i$ masks), 
{\it and} that they are located in the right template, i.e.\ the template 
which scans the right depth range [$x_i$,$x_i$+$\Delta x$] (as they are 
also present in the $M_i$ masks).
This two-step procedure has the advantage to be easy to manage and 
avoids the enormous task of directly identifying which line (among the 
millions) contributes to a given hole before deciding to keep it or not.
Table~\ref{Tab:synth_masks} gives for each atmospheric model the total range 
of reference optical depths $x$ scanned by the masks, the depth step 
$\Delta x$ and the number of lines per mask $C_i$.
The masks $C_i$ are available at 
{\tt http://www-astro.ulb.ac.be/Html/home.html\#tomography}.

% Table Synthetic numerical masks
\begin{table*}
\caption[]{The synthetic templates}
\begin{flushleft}
\begin{tabular}{lccc}
\hline\noalign{\smallskip}
\medskip
Model (\Teff, $\log g$) & 2800~K, $-0.4$ & 3500~K, 0.9 & 4250~K, $-0.5$ \\
\medskip
Total range scanned ($x$-scale) & $-2.75 \rightarrow -6.50$ 
                                & $-2.00 \rightarrow -8.00$ 
                                & $-1.50 \rightarrow -7.50$ \\
$\Delta x$  & 0.75               & 0.75               & 0.75 \\
\hline\noalign{\smallskip}
Masks & \multicolumn{3}{c}{\# of lines per mask} \\
$C_1$ (innermost) & 211 & 777 & 1337 \\ 
$C_2$             & 186 & 610 & 1099 \\
$C_3$             & 193 & 433 &  873 \\
$C_4$             & 160 & 321 &  602 \\
$C_5$             &  99 & 168 &  323 \\
$C_6$             & --- & 167 &  184 \\
$C_7$             & --- &  94 &  124 \\
$C_8$ (outermost) & --- &  46 &   47 \\
\noalign{\smallskip}
\hline
\end{tabular}
\end{flushleft}
\label{Tab:synth_masks}
\end{table*}

At this point, a comment should be made about the validity of the 
  use of {\it static} models to design templates to be used on dynamical 
  atmospheres. Let us first remark that, if one aims at uncovering the
  complex line profiles of LPVs caused by the dynamics of their
  envelope, the correlation of the observed spectrum with a mask
  constructed from a
dynamical model is not the way to go. 
In the ideal case where the dynamical model used to construct the
template   
would perfectly match the real star, the resulting CCF would 
be very sharp and 
single-peaked.
To reveal the various
velocities characterizing the different layers moving in the
envelope, one needs to start instead from templates constructed from
static models!

The real difficulty in the application of `static' templates to
dynamical atmospheres is elsewhere. It is clear that the
contribution functions to the line depression may acquire several local maxima 
in a dynamical atmosphere (see e.g., Albrow \& Cottrell 1996). 
This is precisely why the line shape acquires complex
profiles in dynamical atmospheres, which simply reflect the complex
shape of the 
contribution function. The condition
that needs to be satisfied for the tomographic method to yield
reliable results with `static' masks is the following: 
the contribution functions of {\it all} 
the lines that were assigned to a given static mask 
should change in the {\it same} way when
passing from a static to a dynamic atmosphere. In other words, the
lines supposed to form in the same layers in a static atmosphere
should still form in the same (possibly becoming non-connex) layers in a
dynamical atmosphere. Or equivalently, there should be no swap of lines between
masks when passing from a static to a dynamical atmosphere. We defer
to a forthcoming paper 
the test of how well this assumption is satisfied in practice. 
At this stage, we can only provide empirical, {\it a posteriori} 
evidence
that this condition does not seem to be  very badly violated at least. 
This empirical evidence relies on the fact that (i) 
the CCF shapes vary gently and
smoothly from one mask to the next (see
Figs.~\ref{Fig:ccf_zoph_N1_med} and \ref{Fig:ccf_vtau_N4_med}), 
and (ii) the evolution of the CCF shape from one mask to the next 
reflects what is expected in the framework of 
the Schwarzschild scenario (see Fig.~4 of Paper I).

\subsubsection{Application to non-variable stars}
\label{Sect:nonvar}

\begin{figure*}
%  \resizebox{\hsize}{!}{\rotatebox{90}{\includegraphics{ccf_mgem_N6_3500.ps}}}
  \resizebox{\hsize}{!}{\rotatebox{90}{\includegraphics{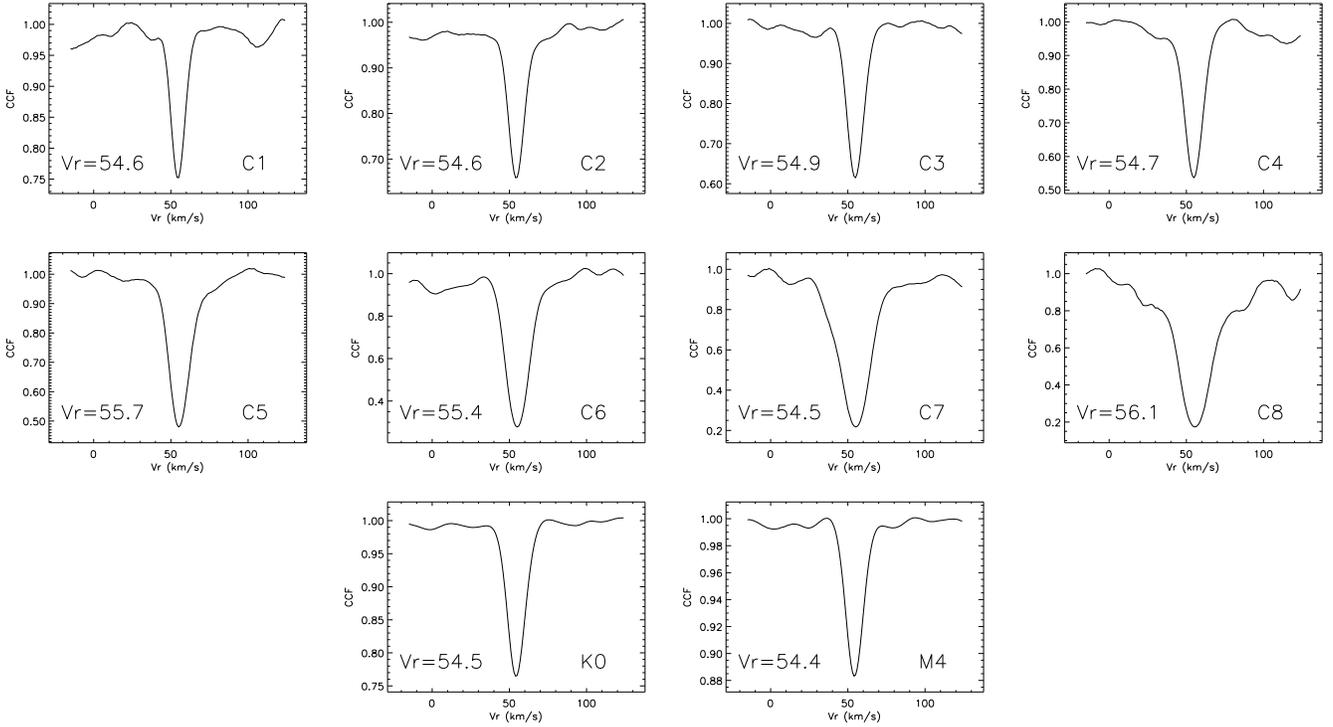}}}
  \caption[]{Sequence of CCFs obtained for $\mu$\,Gem with the synthetic
  templates of the \Teff=3500~K series. Labels refer to the template names 
  (see Table~\ref{Tab:synth_masks}). The sequence goes from the innermost 
  (top left) to the outermost (bottom right) layers. The CCFs 
  obtained with the K0- and the M4-templates are also displayed. Note how
  the CCF wings and contrast increase from $C_1$ to $C_8$}
  \label{Fig:ccf_mgem_N6_med}
\end{figure*}

In order to check whether our synthetic templates are reliable, the CCFs of 
the non-variable stars of the sample have been computed with the newly 
designed masks. The sequence of CCFs obtained for $\mu$~Gem, a M3
giant, with the 3500~K templates is 
presented in Fig.~\ref{Fig:ccf_mgem_N6_med}. The CCFs obtained with the K0- 
and M4-templates are also displayed for comparison.

It can be seen that the synthetic masks $C_i$ yield the same 
radial velocity to within 0.45~\kms\ r.m.s. (excluding the mask C8 which delivers
a very broad CCF). 
Fig.~\ref{Fig:ccf_mgem_N6_med} also shows that the contrast 
of the CCFs increases along the sequence $C_1$ to $C_8$. This is not surprising
since, from $C_1$ to $C_8$, 
each template scans stronger and stronger lines (i.e.\ forming closer and 
closer to the surface), and the CCF contrast reflects the average intensity of
the group 
of lines scanned. It is also remarkable to note that the ``signal-to-noise'' 
ratio of the CCFs (ratio of the fluctuations of the continuum level to the 
depth of the peak) remains more or less constant. Also note how the wings 
of the CCF strengthen from $C_1$ to $C_8$, revealing a similar - and expected
-
change in the profiles of the lines scanned.

A further confirmation of the validity of our method is provided {\it a
posteriori} 
by the
discussion of Sect.~\ref{Sect:individual}, which shows that the CCF correctly
reproduces the complex shapes observed in a dynamical atmosphere for the lines
probed by the considered mask (see Fig.~\ref{Fig:line_ccf}).  

% Figure sequence CCF mu Gem, N6, 3500K -> Fig:ccf_mgem_N6_med

\subsubsection{Application to variable stars}
\label{Sect:var}

\begin{figure*}
%  \resizebox{\hsize}{!}{\rotatebox{90}{\includegraphics{ccf_zoph_N1_3500.ps}}}
  \resizebox{\hsize}{!}{\rotatebox{90}{\includegraphics{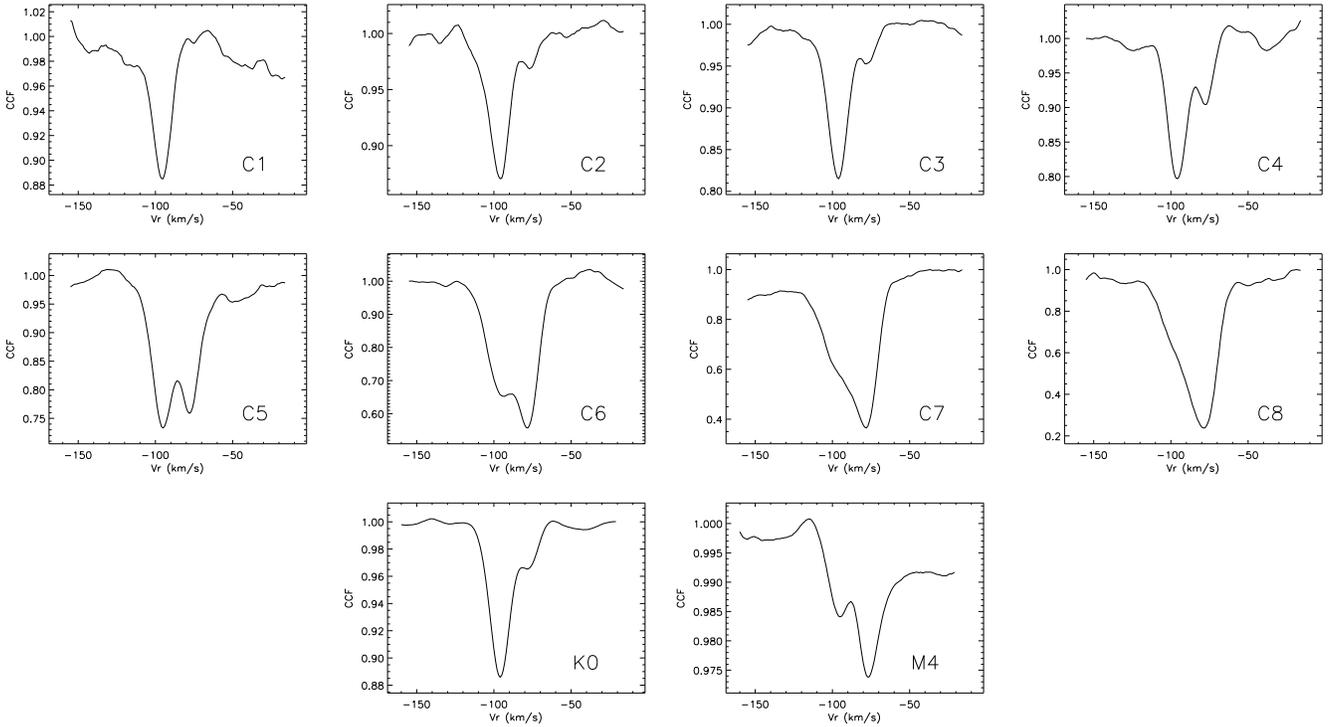}}}
  \caption[]{Sequence of CCFs obtained for Z\,Oph (night N1, phase 0.08) with
the synthetic
  templates of the \Teff=3500~K series. Labels have the same meaning as in
Fig.~\ref{Fig:ccf_mgem_N6_med}}
  \label{Fig:ccf_zoph_N1_med}
\end{figure*}

\begin{figure*}
%  \resizebox{\hsize}{!}{\rotatebox{90}{\includegraphics{ccf_vtau_N4_3500.ps}}}
  \resizebox{\hsize}{!}{\rotatebox{90}{\includegraphics{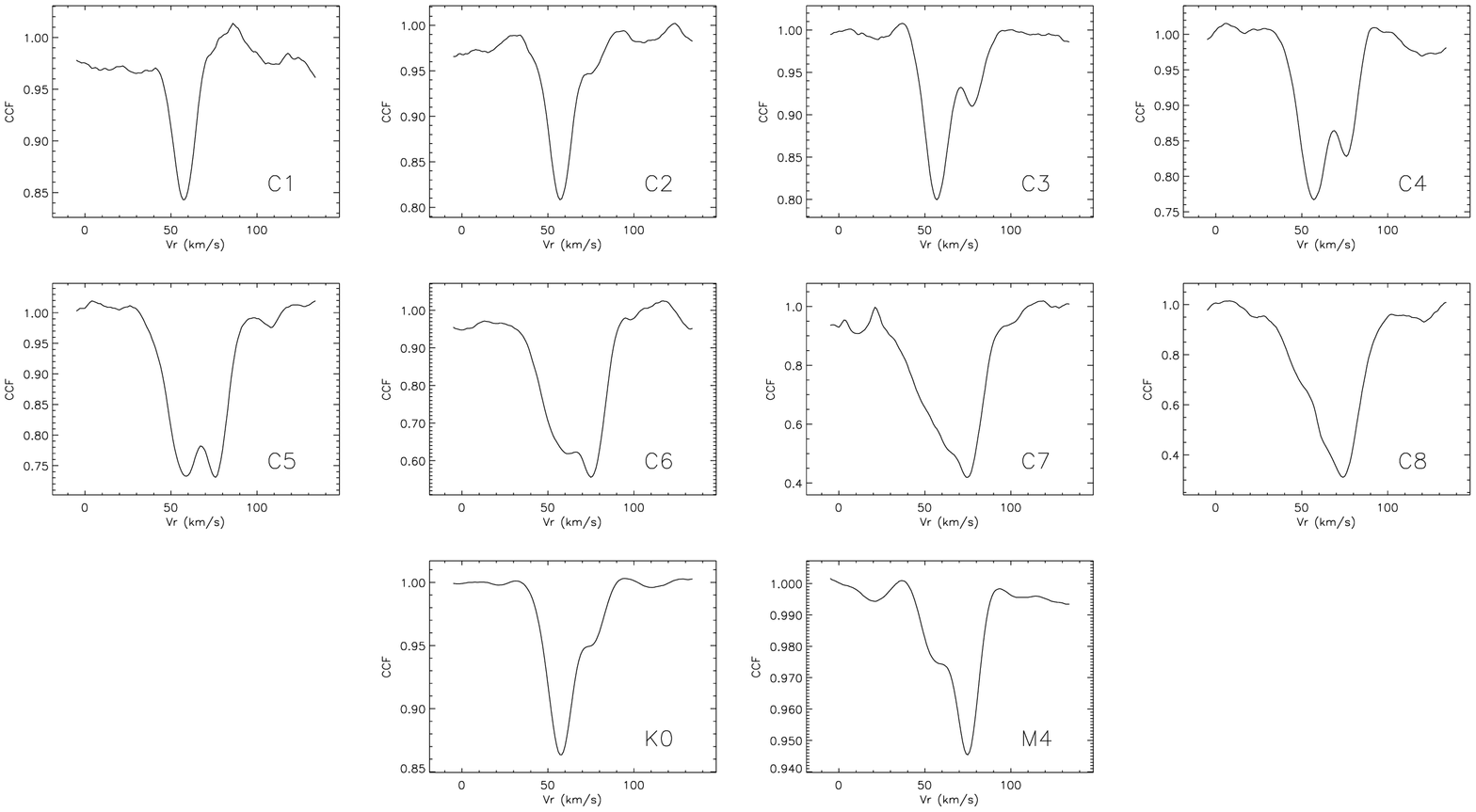}}}
  \caption[]{Sequence of CCFs obtained for V\,Tau (night N4, 
  phase 0.08) with the synthetic
  templates of the \Teff=3500~K series. 
  Labels have the same meaning as in Fig.~\ref{Fig:ccf_mgem_N6_med}}
  \label{Fig:ccf_vtau_N4_med}
\end{figure*}

Now that the reliability of the synthetic masks has been assessed from their
application to non-variable stars (Sect.~\ref{Sect:nonvar}), they 
may be applied to variable stars. Figures~\ref{Fig:ccf_zoph_N1_med} and
\ref{Fig:ccf_vtau_N4_med} present the sequence of CCFs obtained with the 
synthetic templates of the \Teff=3500~K series for Z\,Oph and V\,Tau during
night N1 (phase 0.08) and N4 (phase 0.08) respectively; they clearly reveal 
a transition from a single blue peak (templates $C_1$--$C_3$) to a red one 
($C_8$) through double profiles ($C_4$--$C_7$). Again, the 
\sch\ scenario reveals itself, but this time in a {\it spatial} rather than {\it
temporal} sequence. The deeper the mask scans, the more intense the blue 
peak is (and conversely for the red peak). With the synthetic 
templates, it is thus possible to follow the change of the CCFs from one
layer to the next, i.e.\ to monitor the velocity field across the 
atmosphere.

It is also interesting to note that the CCF obtained with the $C_2$-template
(i.e.\ deep layers) is very similar to the one obtained with the 
K0-template, whereas the $C_6$- and $C_7$-templates (outermost layers) give 
profiles similar to those obtained with the M4-template. This similarity
corroborates the conclusion of
Sect.~\ref{Sect:templates} on the different depths probed by the two default K0-
and M4-templates.

% Figure sequence CCF Z Oph, N1, 3500K -> Fig:ccf_zoph_N1_med

% Figure sequence CCF V Tau, N4, 3500K -> Fig:ccf_vtau_N4_med

\begin{figure*}
%  \resizebox{\hsize}{!}{\rotatebox{90}{\includegraphics{ccf_rtcyg_N3_3500.ps}}}
  \resizebox{\hsize}{!}{\rotatebox{90}{\includegraphics{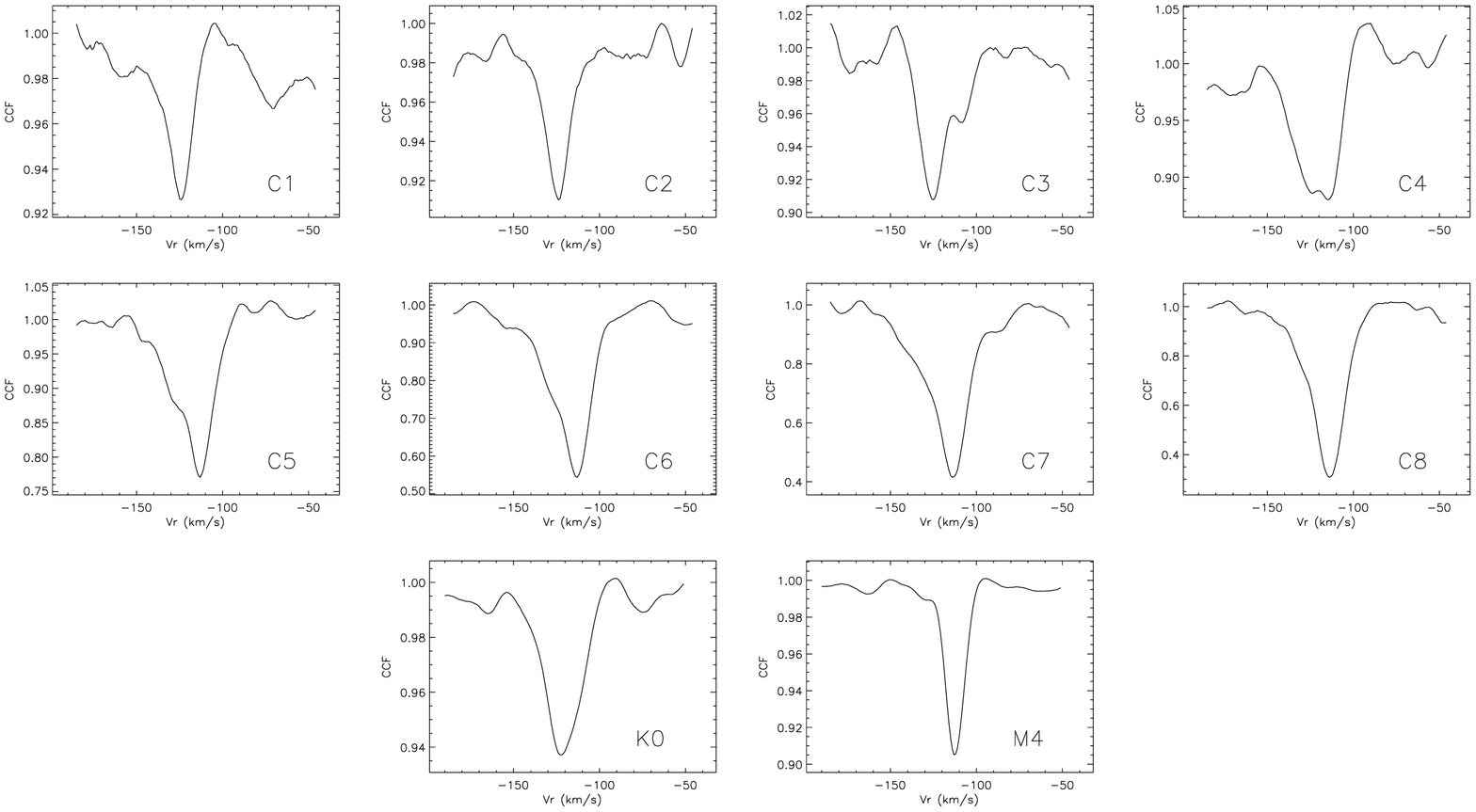}}}
  \caption[]{Sequence of CCFs obtained for RT\,Cyg (night N3, 
  phase 1.23) with the synthetic
  templates of the \Teff=3500~K series.   Labels have the same meaning as in
Fig.~\ref{Fig:ccf_mgem_N6_med}}
  \label{Fig:ccf_rtcyg_N3_med}
\end{figure*}

Fig.~\ref{Fig:ccf_rtcyg_N3_med} provides another illustration of the 
efficiency of the synthetic templates: while each of the default templates 
yields a single peak (although rather asymmetrical for the 
K0-template) for RT~Cyg at phase 1.23, the $C_3$- and $C_4$-templates clearly
reveal that line-doubling is actually present in that star for the specific set
of lines probed by the masks $C_3$ and $C_4$. We refer the reader to Paper~I for
the application of the tomographic masks to a sequence of spectra of RT Cyg
covering phases 0.81 to 1.16, which clearly reveals the outward propagation of
the shock front. 

% Figure sequence CCF RT Cyg, N3, 3500K -> Fig:ccf_rtcyg_N3_med

% Figure sequence CCF RU Her, N10, 3500K -> Fig:ccf_ruher_N10_med
\begin{figure*}
%  \resizebox{\hsize}{!}{\rotatebox{90}{\includegraphics{ccf_ruher_N10_3500.ps}}}
  \resizebox{\hsize}{!}{\rotatebox{90}{\includegraphics{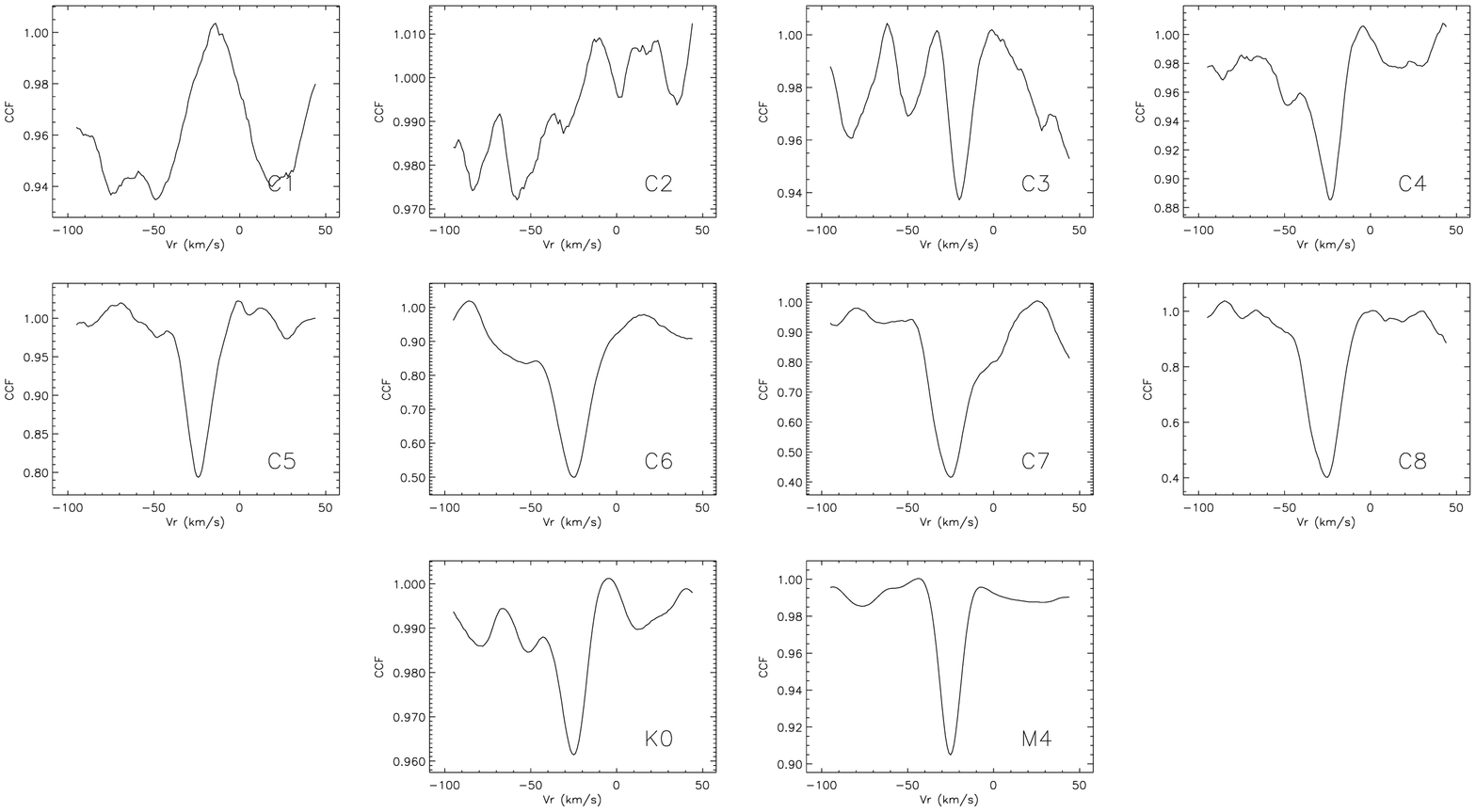}}}
  \caption[]{Sequence of CCFs obtained for the late-type LPV RU\,Her
    (M6-M9; night N10, 
  phase 0.08) with the synthetic
  templates of the \Teff=3500~K series.   Labels have the same meaning as in
Fig.~\ref{Fig:ccf_mgem_N6_med}}
  \label{Fig:latetypeLPV}
\end{figure*}

The tomographic masks 
yield somewhat poorer results for late-type LPVs. 
As an illustration, Fig.~\ref{Fig:latetypeLPV} shows the 
 application of the tomographic
masks to the late-type LPV RU~Her (M6-M9). Like the M4V mask, 
the outer masks C4 
to C8 (of the 3500~K set) yield a
well-contrasted, but sometimes quite asymmetric, CCF. 
The innermost masks C1 and C2 provide a CCF that
is, however, totally useless. 
The C3 mask presents several secondary peaks (that are reminiscent of
those obtained with the K0III template). The peak at $V \sim
50$~\kms\ remains visible in the C4 and C5 CCFs, but it is difficult
to assess whether it traces a real layer or whether it must be
ascribed to some kind of `correlation noise' due to the mismatch
between the template and the observed spectrum. A future 
implementation of the tomographic method (lifting some of the
simplifying hypotheses described in Sect.~\ref{Sect:tomo}) 
should hopefully improve the results for late-type LPVs.

\subsubsection{Individual lines and cross-correlation profiles}
\label{Sect:individual}

\begin{figure*}
%  \resizebox{\hsize}{!}{\includegraphics{compar_line2ccf.ps}}
  \resizebox{\hsize}{!}{\includegraphics{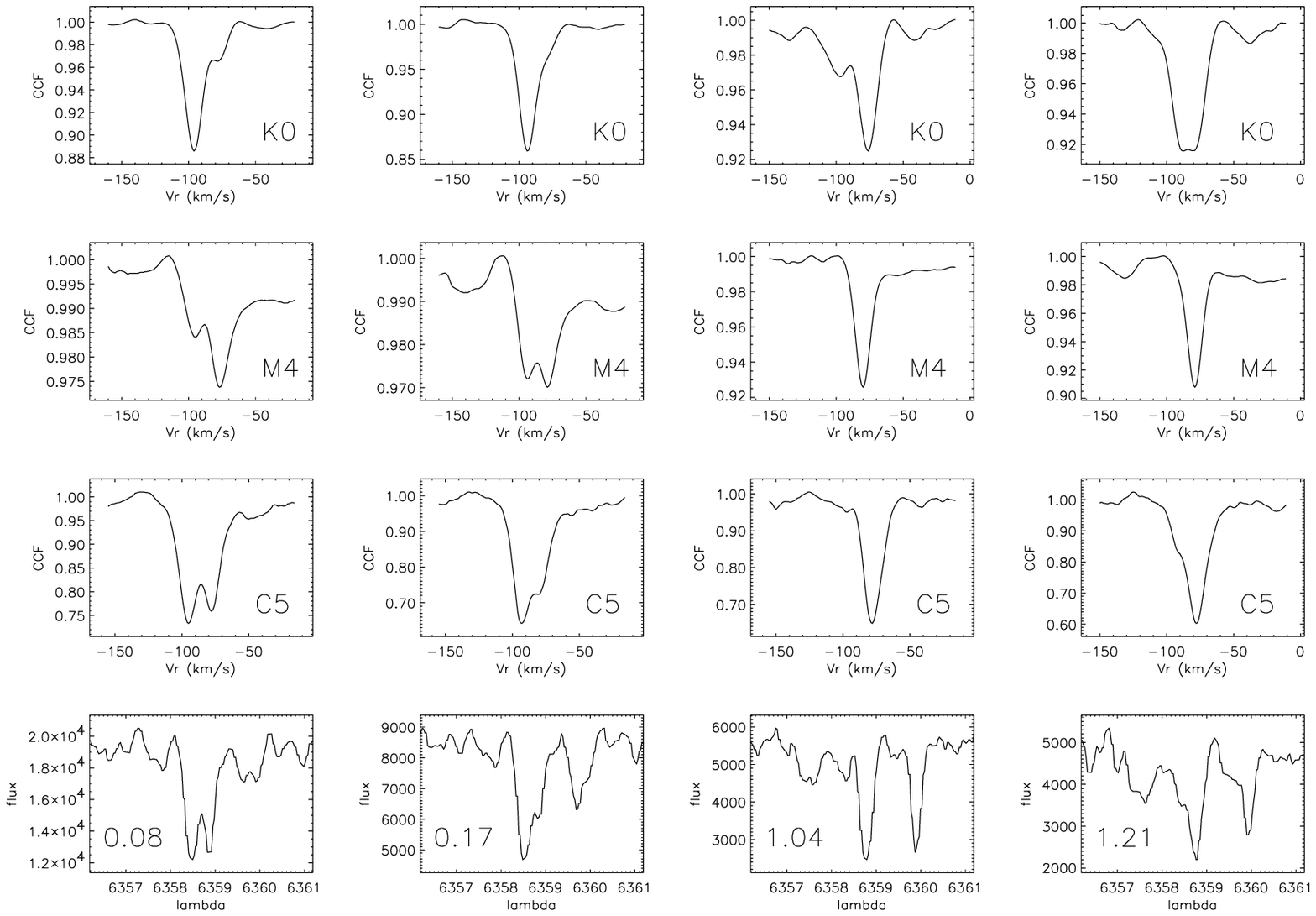}}
  \caption[]{Comparison of the $\lambda$ 6358.69 Fe line as seen in 
  the spectrum of Z\,Oph at phases 0.08, 0.17, 1.04 and 1.21 (bottom row) 
  with the CCFs obtained with the K0-, M4- and $C_5$-templates (from top 
  to bottom)}
  \label{Fig:line_ccf}
\end{figure*}

A definite proof that 
the complex CCF profiles of LPVs as observed for instance  on
Figs.~\ref{Fig:ccf_zoph_N1_med} or
\ref{Fig:ccf_vtau_N4_med} are not an artefact of the method, is 
provided by the comparison of the CCF profile with the shapes of individual
lines probed by the corresponding mask. For this purpose, an early-type LPV must
be selected, to avoid too severe spectral crowding. 
Z\,Oph (K3 at maximum) has the earliest 
spectral type in the sample.
Fig.~\ref{Fig:line_ccf} compares the $\lambda$ 6358.69 Fe 
line at 4 different phases (bottom row) with the CCFs obtained with the K0-, M4-
and $C_5$-templates 
(from top to bottom), since the $\lambda$ 6358.69 line belongs to the
$C_5$-template. The CCFs obtained with this template (third 
row) reproduce remarkably well the shape of the  $\lambda$ 6358.69 line. On the 
contrary, the shapes of the CCFs obtained with the K0- and the M4-templates 
are noticeably different, since they were shown to correspond instead to the
$C_2$ and $C_6-C_7$ templates respectively (Sect.~\ref{Sect:var}). This is
clear evidence that the CCFs obtained 
with the synthetic templates correctly reflect the shape of the 
set of lines they scan (at least in the case of early-type LPVs). 
 This result may be seen as an {\it a
  posteriori}  validation of the tomographic method as a whole
  (at least for early-type LPVs; see Sect.~\ref{Sect:var} and
  Fig.~\ref{Fig:latetypeLPV} for a discussion of
  late-type LPVs)
  and of the
  various assumptions involved (use of `static' masks,
  Eddington-Barbier approximation for the depth of formation of
  spectral lines...) as discussed in Sect.~\ref{Sect:tomo}.

% Figure comparison CCF / individual lines -> Fig:line_ccf

\subsubsection{Comparison of CCFs obtained with synthetic templates constructed
from models of different temperatures}
\label{Sect:coolmasks}
%                                                                           |
This section briefly addresses the question of the influence of the 
model used to construct the synthetic templates on the resulting CCFs of LPVs. 
Browsing the large database containing the CCFs  computed with the tomographic
masks of the \Teff\ = 4250, 3500 and 2800~K series for all 315 available LPV
spectra (Table~3 of Paper~III) 
has shown that (i) the masks based on the 4250~K and 3500~K 
series are the most
appropriate to perform the tomography of LPVs (since the CCFs obtained with
masks $C_1, C_2$ of the 2800~K
series are generally noisy\footnote{note that this poor result may be
  due to the condition $\tau_\lambda = 2/3$ used to define the depth
  of formation of
the line depression, which was shown in Sect.~\ref{Sect:formation} to
be poorly satisfied in the case of faint lines} -- see also Fig.~\ref{Fig:latetypeLPV} in
Sect.~\ref{Sect:var}); 
(ii) no information that is not already present 
in the CCFs of the 4250~K series is obtained with the CCFs from the other two
series. Schematically, the CCFs of the three series match each other after
translating one sequence with respect to the other, i.e., the CCF of template
$C_i$(4250~K) is almost identical to the  CCF of template  $C_{i-1}$(3500~K).

\section{Summary}
\label{Sect:conclusion}

A tomographic method has been described that makes it possible to follow the
propagation of the shock wave across the photosphere of LPV stars. 
The method relies on the 
correlation of the observed spectrum with 
numerical masks probing layers of different atmospheric  depths. The tomographic
masks are constructed from synthetic spectra of red giant stars,
that provide the depth of formation of spectral lines. When applied to
Mira stars around maximum light, they reveal that the deeper layers are
generally characterized by blueshifted absorption lines (translating their
upward motion), whereas the uppermost layers generally exhibit redshifted
absorption lines (translating their infalling motion). A double-peak profile
is then found in intermediate layers, where the shock front is located. At later
phases, the shock front is  seen moving towards upper layers, until it leaves
the photosphere.

A number of checks validating the tomographic method have been presented: (i)
the tomographic masks yield single-peaked CCFs at the same radial velocity (with
a r.m.s. deviation of about 0.5 \kms) when applied to non-pulsating stars; (ii)
the default K0- and M4-templates behave approximately like the $C_2$ and
$C_6-C_7$ templates,
respectively, of the \Teff\ = 3500 K series; (iii) when checked on a  spectrum
of an early-type LPV that is not too crowded, the CCF profile correctly
reproduces the complex shape of the lines belonging to the corresponding mask. 

The application of the tomographic method to a sequence of spectra around
maximum light makes it possible to follow the outward propagation of the shock
front (this was demonstrated in Paper~I for RT Cyg).     

The method offers interesting perspectives for the study of the dynamics of LPV
atmospheres. However, in its current state, its quantitative predictions
(concerning for example the velocity of the shock front) remain rather limited,
mainly because the geometrical radius associated with each mask is currently
unknown. Spectro-interferometric observations would be necessary to make
progress in that direction. Such measurements will probably become feasible with
the new generation of interferometers that will be available in the near future.

\begin{acknowledgements}
RA benefits of a TMR "Marie Curie" Fellowship at ULB. A.J.\ is Research 
Associate from the {\it Fonds National de la Recherche Scientifique} 
(Belgium). 
This program would not have been possible without the generous
allocation of telescope time at the {\it Observatoire de
  Haute-Provence} (operated by C.N.R.S., France).
\end{acknowledgements}

\end{document}